\documentclass[aps,prd,twocolumn,showpacs,showkeys,preprintnumbers,amsmath,amssymb]{revtex4-1}
\usepackage{amsmath}
\usepackage{amssymb}

\begin{document}

\title{On Quasinormal Modes for Gravitational Perturbations of Bardeen Black Hole}
\author{S. C. Ulhoa }
\email{sc.ulhoa@gmail.com} \affiliation{Instituto de F\'{i}sica, Universidade de Bras\'{i}lia, 70910-900, Bras\'{i}lia, DF, Brazil.\\ Faculdade Gama, Universidade de Bras\'{i}lia, 72444-240, Setor Leste (Gama), Bras\'{i}lia, DF, Brazil.}

\begin{abstract}
In this paper we investigate gravitational perturbations of a
regular black hole, particularly Bardeen solution. Such system is
solution of Einstein equations that do not have a singularity at the
origin of the radial symmetry. However it still have events horizons
depending on the values of the characteristic parameters of the
solution. When a black hole is perturbed, it oscillates. It gives
rise to damped vibrating modes which are known as quasinormal modes.
It is calculated the quasinormal frequencies of a regular black hole
using the third order WKB approximation for gravitational
perturbations. The results are presented in tables \ref{tabela0},
\ref{tabela1} and \ref{tabela2}.
\end{abstract}


\maketitle
\section{Introduction}
\noindent

In 1957 Regge and Wheeler explored the stability of Schwarzschild
solution over gravitational perturbations~\cite{PhysRev.108.1063}.
They discovered that it leads to a Schr\"oedinger-like equation with
a specific effective potential and a frequency associated to the
temporal dependence. Since then others perturbations for different
systems have been analyzed and lead to similar
equations~\cite{lrr-1999-2}. The frequencies in those equations can
be calculated by means semi-analytic or numerical methods. Such
frequencies are discrete and complex numbers which can be obtained
by the imposition of specific boundary conditions such as the
existence of ongoing waves at the spatial infinity and ingoing waves
at the event horizon . This is known as quasinormal modes. These
modes are damped oscillations of the spacetime geometry that can be
used to investigate fundamental features of the gravitational field.
The real part is related to the oscillation itself while the
imaginary part refers to the rate at each mode is damped. Therefore
quasinormal modes may be important to gravitational waves physics
and it could become an environment to study the basic effects of
some sort of quantized gravity, since the quasinormal frequencies
are discrete quantities.

Defining a singularity is an intricate task, in what concerns black
holes the presence of a fundamental singularity can cause serious
difficulties when one tries to predict the future from the past,
which led Einstein himself to doubt about the physical realization
of the first solutions of the field equations of general
relativity~\cite{einstein}. It's believed that the singularities
appearing in some solutions, such as Schwarzschild's solution, is
due to highly symmetric assumptions used to get them. However the
singularity is not a privilege of general relativity since the
collapse of a spherical dust in newtonian theory also yields a
singular point. Over the years a great effort has been directed to
deal with such problem by means the development of the so called
cosmic censorship hypotheses~\cite{penrose, Wagh:2002ch}. The
singularity would be inaccessible to an observer outside the event
horizon and always covered by it which allows the avoidance of a
naked singularity. This is a route to contour the problem, not to
solve it, since the singularity, which is a place where no physics
can be done, is still there. Since there is neither a physical
significance nor a experimental support for the presence of a
singularity, one could seek for non-singular solutions of field
equations. In fact, the ideas of gravastars and regular black holes
arose from this kind of feeling. The last objects are interesting
because they do not have a fundamental singularity at the origin of
the coordinate system. Event horizons can be present but they could
be removed under some change of variables.

The first regular solution of Einstein equations that describes a
black hole was due to Bardeen who got his solution only
approximatively~\cite{bardeen}. Later it was discovered that such a
solution could be viewed as the gravitational collapse of a magnetic
monopole to which a nonlinear electromagnetic energy-momentum tensor
works as the source of field equations~\cite{PhysRevLett.80.5056}.
This has lead the Bardeen metric to the category of an exact
solution of Einstein equations.  Others regular solutions was
obtained by the coupling of some matter field, such as scalar field,
to gravitation in a cosmological
context~\cite{Bronnikov:2005gm,PhysRevD.86.024028}. Because of their
singularity-free feature regular black holes could be great
candidates to represent realistic final stages of collapsing regular
configurations such as a common star. Although the success of the
computation of quasinormal modes of black holes or neutron stars, it
has not been done for regular black holes for gravitational
perturbations.

In this paper the quasinormal modes of Bardeen black hole, for
gravitational perturbations, are calculated. Such modes are the
solution of Einstein equations in the presence of a non-linear
electromagnetic field which is an entirely new paradigm when
compared to what is done in \cite{PhysRev.108.1063} for
Schwarzschild black hole, since it is a vacuum solution. However the
first step is to construct axial perturbations once they are simpler
than polar perturbations. It will be chosen a gauge in which there
is no $\phi$ dependence.

The paper is organized as follows.  In section \ref{GP} the general
theory of black holes stability of a spherical symmetric background
metric for gravitational perturbations is summarized. The master
equation and the effective potential are obtained for the Bardeen
solution which was the first one describing a regular black hole. In
section \ref{QNM} the complex frequencies for such a solution has
been calculated. It was done using WKB approximation of third order,
the results are then organized in tables \ref{tabela0},
\ref{tabela1} and \ref{tabela2}. Finally in the last section the
concluding remarks are presented.

\section{Gravitational Perturbations of a Regular Black Hole}\label{GP}
\noindent

In this section we carry out a treatment to deal with a gravitational perturbation of regular black holes. A generic spherically symmetric background metric $\eta_{\mu\nu}$, is defined by the line element

\begin{equation}
ds^2=-f(r)\,dt^2+f^{-1}(r)\,dr^2+r^2(d\theta^2+sin^2\theta
d\phi^2)\,,\label{ds2}
\end{equation}
thus the metric tensor can be written as $g_{\mu\nu}=\eta_{\mu\nu}+h_{\mu\nu}$. The perturbations $h_{\mu\nu}$ can be decomposed as

\begin{equation}
h_{\mu\nu}=
\left(
  \begin{array}{cccc}
    0 & 0 & 0 & h_0(r,t) \\
    0 & 0 & 0 & h_1(r,t) \\
    0 & 0 & 0 & 0 \\
    h_0(r,t) & h_1(r,t) & 0 & 0 \\
  \end{array}
\right)\,h(\theta)\,,\label{1}
\end{equation}
which is similar to the axial decomposition in the Regge-Wheeler gauge~\cite{PhysRev.108.1063}. However it is not the same gauge since we would like to find the form of $h(\theta)$ by means the field equations rather than by imposition of some kind of expansion in terms of spherical harmonics.

The perturbed Einstein equations are given by

$$\delta G_{\mu\nu}=k\delta T_{\mu\nu}\,,$$
where $k=8\pi$ in units such that $G=c=1$ and

\begin{eqnarray}
\delta G_{\mu\nu}&=&\delta R_{\mu\nu}-\frac{1}{2}\left(h_{\mu\nu}R+\eta_{\mu\nu}\delta R\right)\,,\nonumber\\
\delta R&=&\eta^{\mu\alpha}\eta^{\nu\gamma}h_{\alpha\gamma}R_{\mu\nu}\,,\nonumber\\
\delta R_{\mu\nu}&=&-\nabla_\alpha\delta\Gamma^{\alpha}_{\mu\nu}+\nabla_\nu\delta\Gamma^{\alpha}_{\mu\alpha}\,,\nonumber\\
\delta\Gamma^\alpha_{\beta\gamma}&=&\frac{1}{2}\eta^{\alpha\nu}(\partial_{\gamma}h_{\beta\nu}+\partial_{\beta}h_{\gamma\nu}-
\partial_{\nu}h_{\beta\gamma})\,.\nonumber
\end{eqnarray}

Thus, from eq. (\ref{1}), the above equations yield

\begin{small}
\begin{equation*}
\delta G_{23}= \frac{1}{2}\left[-{1\over {f(r)}}{{\partial h_0}\over
{\partial t}}+ {{\partial \lbrack f(r)h_1\rbrack}\over {\partial
r}}\right]\left[\frac{dh(\theta)}{d\theta}-2\frac{\cos\theta}{\sin\theta}\,h(\theta)\right]
\end{equation*}
\end{small}
and
\begin{widetext}
\begin{equation*}
\delta G_{13}=\frac{1}{2}\left\{{1\over {f(r)}}\left[ {{\partial^2h_1}\over {\partial t^2}}-
{{\partial^2h_0}\over {\partial t \partial r}} +{2\over r}
{{\partial h_0}\over {\partial t}} \right]-\left(\frac{2}{r^2}-\frac{2}{r}\frac{df}{dr}-\frac{d^2f}{dr^2}\right)h_1\right\}h(\theta)-
\frac{h_1}{2r^2}\left(\frac{d^2h(\theta)}{d\theta^2}-\frac{\cos\theta}{\sin\theta}\frac{dh(\theta)}{d\theta}\right)\,.\nonumber
\end{equation*}
\end{widetext}
Therefore given a background metric which is a solution of the
unperturbed Einstein equation, it's possible to find the
gravitational perturbations by the above equation once the perturbed
energy-momentum tensor is settled.

\subsection{Master Equation for Bardeen Solution}

There are many works in the literature concerning Bardeen black
hole. Geodesic structure of test particles was studied in
\cite{Zhou:2011aa}, while the features of Bardeen solution as a
gravitational lens was given in \cite{0264-9381-28-8-085008}. The
quasinormal modes of such a spacetime, for scalar field
perturbations, was calculated in \cite{Fernando:2012yw} and a good
revision can be find in \cite{PhysRevD.87.024034}.

Let us consider a non-linear electromagnetic energy-momentum tensor
given by~\cite{PhysRevLett.80.5056}

$$T^\nu_\mu=2\left(L_F\, F_{\mu\lambda}F^{\nu\lambda}-\delta^\nu_\mu\,L\right)\,,$$
where $L_F=\delta L/\delta F$, with
$F=\frac{1}{4}F_{\mu\nu}F^{\mu\nu}$ and
$$L=\frac{3m}{|\alpha|^3}\left(\frac{\sqrt{2\alpha^2F}}{1+\sqrt{2\alpha^2F}}\right)^{5/2}\,.$$
The parameter $\alpha$ represents a magnetic monopole and $m$ its
mass. If we consider
$$F_{\mu\nu}=2\delta^{2}_{[\mu}\delta^{3}_{\nu]}\alpha\sin\theta\,,$$
then we find $F=\alpha^2/2r^4$. Therefore coupling this model of
nonlinear electrodynamics to the Einstein equations, it is possible
to find an exact solution~\cite{bardeen,PhysRevLett.80.5056}. This
solution is known as the Bardeen black hole and it can be put in the
form (\ref{ds2}), with
$$f(r)=1-\frac{2mr^2}{(r^2+\alpha^2)^{3/2}}\,.$$
Such a solution was first obtained only approximately and it was the
first regular black hole ever proposed. We can picture this system
as a self-gravitating magnetic monopole of charge $\alpha$ and mass
$m$. An interesting feature of the Bardeen solution is that it can
have zero, one or two horizons of events depending on the choice of
the magnetic charge $\alpha$. Although it is always a regular black
hole at $r=0$ for $\alpha\neq0$, it describes a regular spacetime
only when the following inequality holds:

$$\alpha^2\leq \frac{16}{27}\,m^2\,.$$ Clearly for $\alpha=0$ the
solution reduces to the well known Schwarzschild metric which do not
represent a regular black hole.

Here we intend to obtain the master equation for gravitational
perturbations of Bardeen spacetime. Thus the non-vanishing
components of the perturbed energy-momentum tensor are

\begin{eqnarray}
\delta T_{03}&=&-2kLh_0(r,t)h(\theta)\,,\nonumber\\
\delta T_{13}&=& -2kLh_1(r,t)h(\theta)\,.\nonumber
\end{eqnarray}

Therefore the perturbed Einstein equations lead to

\begin{eqnarray}
&&\frac{1}{2}\left[-{1\over {f(r)}}{{\partial h_0}\over {\partial t}}+
{{\partial \lbrack f(r)h_1\rbrack}\over {\partial r}}\right]=0\,,\label{h0}\\
&&{1\over {f(r)}}\left[ {{\partial^2h_1}\over {\partial t^2}}-
{{\partial^2h_0}\over {\partial t \partial r}} +{2\over r}
{{\partial h_0}\over {\partial t}} \right]+\nonumber\\
&&+\left[\frac{(\gamma-2)}{r^2}+\frac{2}{r}\frac{df}{dr}+\frac{d^2f}{dr^2}+2kL\right]h_1=0\,,\label{h1}\\
&&\frac{d^2h}{d\theta^2}-\frac{cos\theta}{\sin\theta}\,\frac{dh}{d\theta}+\gamma\,h=0\,.\label{h}
\end{eqnarray}
for the functions $h(\theta)$, $h_0(r,t)$ and $h_1(r,t)$. From the eq. (\ref{h}), it follows that $\gamma=l(l+1)$ and $h(\theta)=P_l(\cos\theta)$ which are the Legendre Polynomials.

If we use the definition $\psi=\left(\frac{1}{r}\right)f(r)h_1(r,t)$, then eqs. (\ref{h0}) and (\ref{h1}) can be combined into a single one which reads

\begin{small}
\begin{eqnarray}
&&\frac{\partial^2\psi}{\partial t^2}-f^2\frac{\partial^2\psi}{\partial r^2}-f\frac{d f}{d r}\frac{\partial\psi}{\partial r}+f\Biggl[\frac{l(l+1)+2(f-1)}{r^2}+\frac{1}{r}\frac{df}{dr}+\nonumber\\
&+&\frac{d^2f}{dr^2}+2kL\Biggr]\psi=0\,.
\end{eqnarray}
\end{small}
In order to put such equation in a more familiar form, we change the variable $r$ to the ``tortoise'' coordinate, defined by $dx=\frac{dr}{f(r)}$, that leads to

\begin{equation}
\frac{\partial^2\psi}{\partial t^2}-\frac{\partial^2\psi}{\partial x^2} + V(x)\,\psi=0\,,
\end{equation}
with the effective potential given by

\begin{small}
$$V=f\left[\frac{l(l+1)+2(f-1)}{r^2}+\frac{1}{r}\frac{df}{dr}+\frac{d^2f}{dr^2}+2kL\right]\,.$$
\end{small}

The temporal dependence is assumed to be given by $\psi=e^{-\imath\omega t}\phi$, thus we have

\begin{equation}
\left[\frac{\partial^2}{\partial x^2}+\omega^2-V(x)\right]\phi(x)=0\,,\label{6}
\end{equation}
which is the master equation for gravitational perturbations of the
Bardeen solution as the background metric. With the solution of this
equation it would be possible to construct the perturbed metric,
then all the features of spacetime would be known. The frequencies
$\omega$ represent dissipative modes in time, thus after the
perturbation the whole black hole will oscillate and then go to a
stable configuration. Such modes oscillate with the so called
quasinormal frequencies.

\section{Quasinormal Modes in WKB Approximation}\label{QNM}
\noindent

The WKB method is a semi-analytic technique used to solve Sch\"oedinger-type equations like eq. (\ref{6}). Once the solution $\phi(x)$ is known it is possible to use two conditions to properly establish it: it is requested that there are only ongoing waves at the infinity and only ingoing waves at the horizon of events. Usually the frequency in WKB approximation is written as~\cite{Konoplya:2011qq}

$$\frac{\imath\,\left(\omega^2-V_0\right)}{\sqrt{-2V_0^{\prime\prime}}}-\sum_{i=2}^{k}\Lambda_i=n+\frac{1}{2}\,,$$
where $V_0$ and $V_0^{\prime\prime}$ are the effective potential and its second derivative respectively, taken at the point of the maximum of $V$. The second term in the left-hand side of the above equation represents the higher order WKB corrections (from 2nd until the k-th order~\cite{PhysRevD.68.024018}). In general a great feature of the WKB method is that it gets better and better as one take the higher orders of approximation (for low values of n and l). As a matter of fact it is a very accurate procedure even when compared to numeric methods~\cite{Konoplya:2011qq}.

In this paper we will work with WKB approximation of third order~\cite{PhysRevD.35.3632} which is given by

\begin{equation}
\omega_{n,l}^2=\left[V_0+(-2V_0^{\prime\prime})^{1/2}\Lambda\right]- \imath\left(n+\frac{1}{2}\right)(-2V_0^{\prime\prime})^{1/2}(1+\Omega)\,,\label{omega}
\end{equation}
where

\begin{widetext}
\begin{equation*}
\Lambda=\frac{1}{(-2V_0^{\prime\prime})^{1/2}}\left[\frac{1}{8}\,\left(\frac{V^{(4)}_0}{V^{\prime\prime}_0}\right)\left(\frac{1}{4}+\beta^2\right)-
\frac{1}{288}\,\left(\frac{V^{\prime\prime\prime}_0}{V^{\prime\prime}_0}\right)^2\left(7+60\beta^2\right)\right]\,,
\end{equation*}
and

\begin{eqnarray}
\Omega&=&\frac{1}{(-2V_0^{\prime\prime})}\Biggl[\frac{5}{6912}\,\left(\frac{V^{\prime\prime\prime}_0}{V^{\prime\prime}_0}\right)^4
\left(77+188\beta^2\right)-\frac{1}{384}\,\left(\frac{(V^{\prime\prime\prime}_0)^2V^{(4)}_0}{(V^{\prime\prime}_0)^3}\right)
\left(51+100\beta^2\right)+\nonumber\\
&+&\frac{1}{2304}\,\left(\frac{V^{(4)}_0}{V^{\prime\prime}_0}\right)^2\left(67+68\beta^2\right)
+\frac{1}{288}\,\left(\frac{V^{\prime\prime\prime}_0V^{(5)}_0}{(V^{\prime\prime}_0)^2}\right)\left(19+28\beta^2\right)
-\frac{1}{288}\,\left(\frac{V^{(6)}_0}{V^{\prime\prime}_0}\right)\left(5+4\beta^2\right)\Biggr],\nonumber
\end{eqnarray}
\end{widetext}
Here $\beta=n+\frac{1}{2}$ and $V^{(n)}_0=\frac{d^n V}{dx^n}|_{x=x_0}$. The point $x_0$ is the solution of the equation  $\frac{dV}{dx}(x_0)=0$, which means that it is a point of maximum of the effective potential. It should be noted that $x$ is the well known tortoise coordinate.

Using expression (\ref{omega}) it is possible to calculate the quasinormal frequencies which are presented in tables \ref{tabela0}, \ref{tabela1} and \ref{tabela2} for Bardeen space-time. All results refer to quasinormal modes for gravitational perturbations. They were calculated in units of m which means $\omega_{nl}=(m\,\omega_{n,l})$, where $\omega_{nl}$ is dimensionless.

\begin{widetext}

\begin{table}
\caption{ Quasinormal modes of Bardeen space-time for $\alpha=0$\,.}
        \begin{tabular}{ccccccccccc}
    \hline\hline
    $l$ & $$ & $n$ & $$ & $\omega_{nl}$& $$ & $l$ & $$ & $n$ & $$ & $\omega_{nl}$ \\
    \hline
    $1$&$$&$0$&$$&$0.1171192993-0.08879106527i$&$$&$4$&$$&$0$&$$&$0.8090978140-0.09417105983i$\\
    $$&$$&$1$&$$&$0.05501430875-0.2873057745i$&$$&$$&$$&$1$&$$&$0.7964989017-0.2843663754i$\\
    $2$&$$&$0$&$$&$0.3731620888-0.08921749033i$&$$&$$&$$&$2$&$$&$0.7736360348-0.4789739946i$\\
    $$&$$&$1$&$$&$0.3460174754-0.2749155289i$&$$&$$&$$&$3$&$$&$0.7433125214-0.6783003279i$\\
    $$&$$&$2$&$$&$0.3029353684-0.4710642944i$&$$&$5$&$$&$0$&$$&$1.012252026-0.09487331582i$\\
    $3$&$$&$0$&$$&$0.5992651163-0.09272839457i$&$$&$$&$$&$1$&$$&$1.002148953-0.2858304756i$\\
    $$&$$&$1$&$$&$0.5823546522-0.2814060077i$&$$&$$&$$&$2$&$$&$0.9832631880-0.4798984691i$\\
    $$&$$&$2$&$$&$0.5531999534-0.4766840022i$&$$&$$&$$&$3$&$$&$0.9574780375-0.6777975349i$\\
    $$&$$&$3$&$$&$0.5157471801-0.6774290938i$&$$&$$&$$&$4$&$$&$0.9263580230-0.8791948073i$\\
    \hline\hline
    \label{tabela0}
  \end{tabular}
\end{table}

\begin{table}
\caption{Quasinormal modes of Bardeen space-time for $\alpha=0.3$\,.}
        \begin{tabular}{ccccccccccc}
    \hline\hline
    $l$ & $$ & $n$ & $$ & $\omega_{nl}$& $$ & $l$ & $$ & $n$ & $$ & $\omega_{nl}$ \\
    \hline
    $0$&$$&$0$&$$&$0.05982992424-0.3650702720i$&$$&$3$&$$&$0$&$$&$0.6267710057-0.09160068738i$\\
    $$&$$&$$&$$&$$&$$&$$&$$&$1$&$$&$0.6128131830-0.2776876367i$\\
    $1$&$$&$0$&$$&$0.1884164237-0.2186679687i$&$$&$$&$$&$2$&$$&$0.5891039911-0.4697416379i$\\
    $$&$$&$1$&$$&$0.6363972564-0.7536092924i$&$$&$$&$$&$3$&$$&$0.5593512286-0.6667476229i$\\
    $$&$$&$2$&$$&$1.329439135-1.493839029i$&$$&$4$&$$&$0$&$$&$0.8355064868-0.09308120018i$\\
    $2$&$$&$0$&$$&$0.4066400107-0.08797961013i$&$$&$$&$$&$1$&$$&$0.8244154310-0.2809325272i$\\
    $$&$$&$1$&$$&$0.3918842187-0.2699024702i$&$$&$$&$$&$2$&$$&$0.8043475445-0.4728380035i$\\
    $$&$$&$2$&$$&$0.3720498716-0.4598638795i$&$$&$$&$$&$3$&$$&$0.7779038150-0.6691359619i$\\
    $$&$$&$3$&$$&$0.3501499693-0.6529380130i$&$$&$$&$$&$4$&$$&$0.7467267888-0.8688420648i$\\
    \hline\hline
    \label{tabela1}
  \end{tabular}
\end{table}

\begin{table}
\caption{Quasinormal modes of Bardeen space-time for $\alpha=\frac{4}{\sqrt{27}}$\,.}
        \begin{tabular}{ccccccccccc}
    \hline\hline
    $l$ & $$ & $n$ & $$ & $\omega_{nl}$& $$ & $l$ & $$ & $n$ & $$ & $\omega_{nl}$ \\
    \hline
    $0$&$$&$0$&$$&$0.6218228066-0.07896506498i$&$$&$3$&$$&$0$&$$&$0.9315004445-0.07399275203i$\\
    $$&$$&$$&$$&$$&$$&$$&$$&$1$&$$&$0.9189649894-0.2227244676i$\\
    $1$&$$&$0$&$$&$0.6767694809-0.07815597820i$&$$&$$&$$&$2$&$$&$0.8943294277-0.3736175009i$\\
    $$&$$&$1$&$$&$0.6619780155-0.2360238599i$&$$&$$&$$&$3$&$$&$0.8584153696-0.5278493353i$\\
    $$&$$&$2$&$$&$0.6340480798-0.3978834282i$&$$&$4$&$$&$0$&$$&$1.109517971-0.07301443485i$\\
    $2$&$$&$0$&$$&$0.7828924561-0.07622286290i$&$$&$$&$$&$1$&$$&$1.099525117-0.2193579532i$\\
    $$&$$&$1$&$$&$0.7688856410-0.2299687824i$&$$&$$&$$&$2$&$$&$1.079605666-0.3666520402i$\\
    $$&$$&$2$&$$&$0.7420488856-0.3872459558i$&$$&$$&$$&$3$&$$&$1.049912024-0.5155433322i$\\
    $$&$$&$3$&$$&$0.7042736801-0.5493249063i$&$$&$$&$$&$4$&$$&$1.010709699-0.6666705011$\\
\hline\hline
  \end{tabular}
\label{tabela2}
\end{table}
\end{widetext}

Firstly we note that the Bardeen spacetime for $\alpha=0$ is exactly
the Schwarzschild spacetime. It's well known that the ground state
of the quasinormal modes for Schwarzschild metric is given by
$\{l,n\}=\{2,0\}$, thus modes which are below this state have no
physical meaning, although they are charted on table \ref{tabela0}.
It also interesting to note that the imaginary part of the frequency
which is the damping term is very similar for each value of $\alpha$
presented in the tables above. However in the table \ref{tabela1} we
see a peculiar behavior for the real part of the frequency, for
$l=1$ it increases when compared to each value of n. The general
tendency is an increasing of the imaginary part and a decreasing of
the real part of the frequency. Therefore, guided by the ground
state of the quasinormal modes of Schwarzschild spacetime, it seems
natural to also consider the ground state as $\{l,n\}=\{2,0\}$ on
tables \ref{tabela1} and \ref{tabela2}.

\section{Conclusion}
In this article we have analyzed gravitational perturbations of a
regular black hole, by calculating the quasinormal frequencies. Such
results were obtained for the Bardeen solution of Einstein equations
that represent a regular black hole. To obtain the quasinormal modes
it has been used third order WKB approximation which yields good
results when compared to accurate numerical technics, for low values
of the numbers n and l. It is such a surprise to find that the
function $h(\theta)$ is equal to the Legendre Polynomials since that
in the Regge-Wheeler gauge the ``$\theta$'' dependence is given by
the function $h(\theta)=\sin\theta\partial_{\theta}P_l(\cos\theta)$.
However in the Regge-Wheeler gauge it appears the value $l(l+1)$ in
the radial equation which is the same quantity found here due to the
eigenvalue of Legendre equation. The results are summarized in
tables \ref{tabela0}, \ref{tabela1} and \ref{tabela2}. The
quasinormal modes for Schwarzschild were already known, for instance
see table III of reference ~\cite{PhysRevD.35.3632}, however they
are charted here to serve as a comparison guide.


\end{document}